\documentclass[10pt,journal,final,twocolumn]{IEEEtran} 

\usepackage{epsfig}
\usepackage{cite}
\usepackage{tabularx}
\usepackage{amsmath,amssymb,amsfonts,bm}
\usepackage{graphicx}
\usepackage{textcomp}
\usepackage{array,multirow,color,booktabs,textcomp}
\usepackage{float}
\usepackage[dvipsnames, svgnames, x11names]{xcolor}
\usepackage{enumitem}

\newcommand{\FigScal}{1}        
 
\newcommand{\FIGWidtF}{0.8\columnwidth}
\newcommand{\FIGWidtS}{\columnwidth}
\newcommand{\FIGWidtT}{\columnwidth}
\newcommand{\FIGWidto}{\columnwidth}

\newcommand{\TabWidt}{\columnwidth}
\newcommand{\TabCWidF}{3cm}
\newcommand{\TabCWidS}{2.5cm}
\newcommand{\TabCWidT}{2.3cm}
\newcommand{\TabCWidTa}{2.1cm}
\newcommand{\TabCWidu}{3.5cm}
\newcommand{\TabCWidi}{1.4cm}
\newcommand{\TabCWidx}{0.9cm}
\newcommand{\TabCWidxa}{1.9cm}
\newcommand{\TabCWidxb}{1.4cm}

\begin{document}

\title{Unsupervised Face-Masked Speech Enhancement Using Generative Adversarial Networks With Human-in-the-Loop Assessment Metrics}
%
\author{
\IEEEauthorblockN{Syu-Siang Wang, Jia-Yang Chen, Bo-Ren Bai, Shih-Hau Fang \textit{Senior IEEE Member} and Yu Tsao \textit{Senior IEEE Member}}}

%
%
%
%
%

\maketitle

\begin{abstract}
The utilization of face masks is an essential healthcare measure, particularly during times of pandemics, yet it can present challenges in communication in our daily lives. To address this problem, we propose a novel approach known as the human-in-the-loop StarGAN (HL--StarGAN) face-masked speech enhancement method. HL--StarGAN comprises discriminator, classifier, metric assessment predictor, and generator that leverages an attention mechanism. The metric assessment predictor, referred to as MaskQSS, incorporates human participants in its development and serves as a ``human-in-the-loop" module during the learning process of HL--StarGAN. The overall HL--StarGAN model was trained using an unsupervised learning strategy that simultaneously focuses on the reconstruction of the original clean speech and the optimization of human perception. To implement HL--StarGAN, we created a face-masked speech database named ``FMVD," which comprises recordings from 34 speakers in three distinct face-masked scenarios and a clean condition. We conducted subjective and objective tests on the proposed HL--StarGAN using this database. The outcomes of the test results are as follows: (1) MaskQSS successfully predicted the quality scores of face-masked voices, outperforming several existing speech assessment methods. (2) The integration of the MaskQSS predictor enhanced the ability of HL--StarGAN to transform face-masked voices into high-quality speech; this enhancement is evident in both objective and subjective tests, outperforming conventional StarGAN and CycleGAN-based systems.
\end{abstract}

\section{Introduction}
\label{sec:intro}
\IEEEPARstart{T}{he} utilization of masks is an effective healthcare measure, particularly during pandemics, as they play a vital role in comprehensive infection prevention strategies within healthcare systems \cite{wang2021association}. However, the act of wearing masks can lead to speech distortion, creating communication challenges between individuals \cite{alkharabsheh2022effect}. This can significantly impact interactions between healthcare professionals and patients \cite{atcherson2021acoustic, yi2021adverse}. Typically, common face masks, such as N95, cotton masks, and plastic shields act as low-pass filters, primarily dampening the higher-frequency components of speech above 7 \textit{k}Hz \cite{poon2022communication,mallol2021filtering, aliabadi2022influence}. Consequently, face-masked speech can be considered a type of distorted speech, with characteristics differing from those of noise-corrupted and reverberant speech in normal environments. To address this distortion effect, the implementation of a crucial front-end speech-processing technique, known as speech enhancement (SE) is essential. SE aims to restore clean speech from distorted input, thereby enhancing sound quality and intelligibility while improving performance for downstream applications \cite{kang1998speech}.

Generally, an SE system utilizes a mapping function to transform distorted speech into enhanced speech. Various methods to perform SE tasks have been proposed, with the most common being used to handle distortions caused by additive noise \cite{mohammadiha2013supervised,kim2018unpaired,neri2021unsupervised,venkataramani2019style}. These methods aim to reduce noise components from noisy inputs to restore clean speech. Conventional SE methods utilize noise-tracking functions, either explicitly or implicitly, to estimate noise components within the noisy input speech \cite{zheng2023sixty}. Subsequently, a gain function is derived to filter the estimated noise components during the enhancement phase. These gain functions are derived assuming that speech and noise signals are mutually independent \cite{xu2021speech}. Well-known conventional SE methods include spectral subtraction \cite{boll1979suppression} and Wiener filter \cite{lim1979enhancement} with various extensions \cite {lu2008geometric, chen2008fundamentals}. Additionally, certain SE approaches were developed based on assumed probabilistic models of speech and noise signals. Notable examples include the minimum mean-square-error estimator \cite{ephraim1984speech}, maximum a posteriori spectral amplitude estimator \cite{lotter2005speech}, and maximum likelihood spectral amplitude estimator \cite{mcaulay1980speech}. Although these conventional SE approaches typically yield satisfactory enhancement results in stationary or quasi-stationary noise conditions, they may underperform when confronted with non-stationary noises, where the assumed statistical properties of speech and noise deviate from the application scenarios.

Recently, deep learning (DL) models have been leveraged for SE techniques \cite{sekiguchi2019semi, hu20g_interspeech}. These DL-based SE techniques learn the mapping function using a data-driven approach, without relying on the predefined properties of speech and noise signals \cite{DDAE, lee2018speech, speakerddae}. Depending on the availability of paired distorted–clean training data, DL-based SE can be categorized into supervised and unsupervised learning methods. Generally, supervised-learning-based SE outperforms unsupervised-learning-based approaches. However, obtaining paired distorted, clean training data may not always be feasible in real-world scenarios. In contrast, unsupervised learning-based SE methods can be trained without paired distorted–clean data. In practice, the unsupervised learning-based SE methods can serve as a decent initial model that can be fine-tuned to particular tasks, when paired training data become available. The unsupervised learning-based SE approaches can be classified into three categories. The first category involves training the SE model using only noisy speech with noisy-target training (NyTT) \cite{fujimura2021noisy} being a representative method in this category. NyTT involves mixing the original noisy speech with extraneous noise to create an even noisier input that requires enhancement. The SE model is then trained to recover the original noisy speech from this noisier input. The second category involves building an SE model using only clean speech data. A previous study, \cite{bie2022unsupervised}, exploited an unsupervised learning algorithm on a clean corpus to train a dynamical variational autoencoder (VAE). Given the noisy input, the model was used to generate a robust acoustic representation, which eventually resulted in an enhanced voice. In another study, a vector-quantized VAE was trained using clean speech data, incorporating a novel self-distillation mechanism and adversarial training to train the SE model in a self-supervised manner \cite{fu2024self}. The third category of approaches involves estimating SE models using unpaired distorted–clean training data \cite{xiang2020parallel, meng2018cycle, yu2021two,yu2021cyclegan}, utilizing the cycle-consistent learning framework initially developed for image translation \cite{zhu2017unpaired} and adapted for speech processing applications, such as voice conversion (VC) \cite{kaneko2019cyclegan}. Paired generator and discriminator model architectures were used when implementing a cycle-consistent generative adversarial network (CycleGAN) to implement an unsupervised SE system. The generator system converts the speech from noisy to clean, whereas the discriminator function distinguishes between real samples (real noisy and clean speech) and fake samples forged by a generator. While training a cycle-consistent-based SE network, estimating an accurate clean-to-noisy generator is challenging owing to the lack of specified target noise to be generated, limiting the achievable enhancement performance. To address this limitation, a previous study introduced conditional CycleGAN to specify the target noise during training to perform SE \cite{ting2022speech}. In this conditional cycle-consistent-based neural model, the target domain identity serves as an auxiliary feature and is positioned at the input side of the generator and discriminator. This feature guides the model to generate outputs that satisfy the target noise or clean distribution. Owing to its effective interpretive ability and performance, the conditional CycleGAN model structure has proven to be highly effective and versatile, finding applications in various fields, such as VC \cite{lee2020many} and image processing \cite{almahairi2018augmented}.

Enhancing noisy speech using DL-based methods typically results in high-quality output. However, the optimal denoising performance is often achieved when the SE system is trained and tested in the same noisy environment \cite{gonzalez2023assessing,yu2020speech}. In addition, extensive databases with diverse noise conditions are required to achieve optimal performance. Consequently, implicit constraints embedded within DL models can limit the SE capability of a system in specific noisy environments, even if these conditions are part of the training set. This limitation, caused by variations in noise types, has been documented in previous studies \cite{bengio2009curriculum,kumar2010self,chang2017active}. This study focuses on enhancing the sound quality of face-masked speech rather than simply reducing noise. Therefore, standard SE methods that focus on noise reduction may not be directly applicable to the face-masked SE tasks addressed in this study.

The mean opinion score (MOS) is a widely used metric for assessing the performance of SE systems. The MOS scores range from 1 to 5, with higher scores indicating better quality. However, obtaining an unbiased MOS assessment requires recruiting a sufficient number of participants to listen to multiple utterances, rendering subjective evaluations time-consuming and costly. Consequently, objective evaluation metrics have been developed as alternatives to human listening tests to quantify specific characteristics of speech signals. These metrics include the scale-invariant signal-to-distortion ratio \cite{ma2020optimal}, perceptual evaluation of speech quality (PESQ) \cite{PESQ}, and perceptual objective listening quality analysis \cite{POLQA}.

Recently, DL techniques have been widely utilized to build models that approximate human listening evaluations \cite{dong2020attention,kumar2023torchaudio,mittag2021nisqa,cooper2022generalization}. For example, MOSNet \cite{lo2019mosnet} is an end-to-end evaluation model that assesses the naturalness of speech generated by VC or text-to-speech (TTS) system \cite{xu2022deep,maiti2023speechlmscore}. Additionally, DNSMOS \cite{kareddy2021interspeech,DNSMOS} utilized a convolutional neural network (CNN) framework along with ITU-T P.808 metrics to accurately estimate MOS scales. InQSS \cite{chen2021inqss} is a multi-task learning framework that predicts both the PESQ and short-term objective intelligibility (STOI) \cite{taal2011algorithm} scores from spectrogram and scatter coefficient \cite{mallat2012group} inputs. Moreover, these DL-based speech evaluation metrics have been incorporated into various SE studies to enhance the generalization and performance of enhancement systems, ultimately aiming to provide a superior listening experience. For example, works in \cite{MetricGAN, fu2022metricgan} have combined metric prediction networks and discriminators to enhance SE performance.

In this study, we introduce the innovative human-in-the-loop StarGAN (HL--StarGAN) face-masked SE method, designed to address speech distortions caused by wearing masks. The HL--StarGAN model architecture is based on the StarGAN framework \cite{kameoka2018stargan, choi2018stargan}, which comprises a generator, classifier, and discriminator. By integrating conditional cycle-consistent (CC) learning algorithms, the HL--StarGAN method leverages target attribute vectors across all functions to enhance the overall system performance. The classifier embedded within the StarGAN system serves the purpose of categorizing the output generated by the generator. In our implementation, the HL--StarGAN model s composed of a generator, classifier, discriminator, and metric prediction network, with the introduction of a novel ``MaskQSS" as the metric prediction network. The development of the MaskQSS model involves human participants, establishing it as a ``human-in-the-loop" module to enhance HL--StarGAN and is utilized for evaluating voice quality under face-masked scenarios. In particular, we developed a mask-oriented generation system in which the generator processes input speech and a target condition attribute (indicating a mask type or clean) to generate the desired speech output. The resulting output is then assessed using a classifier to determine the type of mask, and a metric prediction network to compute a perceptual score. To construct the metric prediction network, we compile a database of recordings from 34 speakers wearing three types of masks (N95, cotton, and plastic shields), in addition to clean speech recordings. From this database, a subset of face-masked recordings was selected for subjective testing to obtain MOS quality ratings. These paired MOS ratings and corresponding recordings were utilized to train MaskQSS. Simultaneously, the HL--StarGAN system was trained in an unsupervised manner owing to the unavailability of pair-wise face-mask-free clean speech signals. In our experiments, we evaluated the proposed MaskQSS model and HL--StarGAN SE system. The results demonstrated a high Pearson correlation coefficient (PCC) of 91\% between the index scores predicted by MaskQSS and the subjective MOS score, surpassing that of MOSNet at 45\%. Additionally, HL--StarGAN-enhanced speech demonstrated favorable DNSMOS and MaskQSS scores, validating its superior face-masked SE performance compared with that of the baseline StarGAN method.

The contributions of our proposed method are fourfold. (1) The HL--StarGAN framework represents a novel face-masked SE method that aims to enhance the influence of speech in face-masked scenarios rather than reduce general noise in noisy speech. (2) We created and released an FMVD database, which includes clean speech and voice recordings under three distinct mask conditions: ``N95," ``Cotton," and ``Plastic Shield." (3) A novel MaskQSS model was proposed and proven effective in predicting the quality of face-masked speech. (4) We demonstrated the effectiveness of the proposed HL--StarGAN in generating high-quality voices by incorporating ``human feedback" during the model learning phase. To the best of our knowledge, this study is the first attempt to explore an assessment model for face-masked speech, while utilizing the model for face-masked SE. The remainder of this paper is organized as follows. Section \ref{sec:rw} introduces the StarGAN and InQSS models. Section \ref{sec:ccafv} examines the proposed HL--StarGAN SE framework. Section \ref{sec:ea} discusses the experimental setup and an analysis of the results. Finally, Section \ref{sec:summary} summarizes our key findings.

\section{Related studies}\label{sec:rw}
This section delves into the structure of the StarGAN model and the InQSS metric prediction framework.

\subsection{StarGAN}
\begin{figure}[!tp]
    \centering
    \includegraphics[scale=\FigScal]{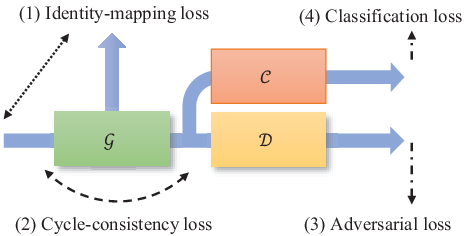}
    \caption{Block diagram of StarGAN, comprising a generator $\mathcal{G}$, discriminator $\mathcal{D}$, and domain classifier $\mathcal{C}$. The total loss, which is a combination of four losses, is utilized to train the generator.}
    \label{fig:stargan}
\end{figure}

As shown in Fig. \ref{fig:stargan}, the StarGAN model comprises three key components: a generator $\mathcal{G}$, discriminator $\mathcal{D}$, and domain classifier $\mathcal{C}$. The generator module is composed of an encoder and a decoder, which convert the input source speech (along with the corresponding spectral feature $\mathbf{Y}$) into a latent representation. The latent representation is then concatenated with the target domain attribute, represented by a one-hot vector $\mathbf{t}$, where non-zero elements indicate the desired identities (such as speaker IDs in VC tasks). Utilizing this concatenated vector, the generator produces the spectral feature $\tilde{\mathbf{X}}$ of the target domain via: $\tilde{\mathbf{X}}=\mathcal{G}\lbrace\mathbf{Y},\mathbf{t}\rbrace$. The discriminator and classifier modules regulate and enhance the performance of the generator, thereby improving the overall effectiveness of the unsupervised learning. During the training phase, the StarGAN system utilizes adversarial training, identity mapping, classification, and CC loss functions, as shown in Fig. \ref{fig:stargan}.

\subsubsection{Adversarial loss function}
The adversarial loss function defined in Eq. \eqref{eq:adv} is minimized to update the $\mathcal{G}$ and $\mathcal{D}$ networks. By optimizing $\mathcal{L}_{adv}^{\mathcal{D}}$, the discriminator can differentiate between real speech recordings $\mathbf{X}$ and the face-mask-enhanced speech $\tilde{\mathbf{X}}$ generated by the generator. The discriminator output is a scalar, with one indicating real recordings and zero representing fake speech. Furthermore, by minimizing $\mathcal{L}_{adv}^{\mathcal{G}}$, the speech generated by the generator in the target domain is expected to deceive the discriminator.
    \begin{equation}\label{eq:adv}
			\begin{aligned}
                \mathcal{L}_{adv}^{\mathcal{D}}&=-E\{\ln[\tfrac{\mathcal{D}\{\mathbf{X}\}}{\mathcal{D}\{\mathcal{G}\{\mathbf{Y},\mathbf{t}\}\}}]\},\\
                \mathcal{L}_{adv}^{\mathcal{G}}&=-E\{\ln[\mathcal{D}\{\mathcal{G}\{\mathbf{Y},\mathbf{t}\}\}]\},
			\end{aligned}
	\end{equation}
    where ``$E\{\cdot\}$'' represents the expectation operation. 
	
	\subsubsection{Classification loss function}
	To determine the type of face mask of a given input, classifier $\mathcal{C}$ is learned in terms of loss function, $\mathcal{L}_{cls}^{\mathcal{C}}$, which is formulated as follows:
    \begin{equation}\label{eq:cls}
			\begin{aligned}
				\mathcal{L}_{cls}^{\mathcal{C}}&=-{E}\{\ln[ p_\mathcal{C}(\mathcal{C}\{\mathbf{X}\})]\},\\
				\mathcal{L}_{cls}^{\mathcal{G}}&=-{E}\{\ln[ p_\mathcal{C}(\mathcal{C}\{\mathcal{G}\{\mathbf{Y},\mathbf{t}\}\})]\}.
			\end{aligned}
	\end{equation}
	Furthermore, from the equation, the generator is updated by minimizing $\mathcal{L}_{cls}^{\mathcal{G}}$ to provide a voice under a specific face-masked scenario.
	
	\subsubsection{Cycle-consistency loss function}
 Equation \eqref{eq:cyc} represents the cycle-consistency loss that enables the generator to generate speech based on specific attributes.
	\begin{equation}\label{eq:cyc}
			\begin{aligned}
                \mathcal{L}_{cyc}^{\mathcal{G}}={E}\{\| \mathcal{G}\{\mathcal{G}\{\mathbf{Y}, \mathbf{t}_c\},\mathbf{t}_f\}-\mathbf{Y}\|_1\},
			\end{aligned}
	\end{equation}
   where $\mathbf{t}_c$ and $\mathbf{t}_f$ represent the clean and face-masked attributes, respectively. With Eq. \eqref{eq:cyc}, given an input speech, the generator can replace the original background or environment with the desired target background, thereby producing a corresponding output voice based on the provided condition attributes.
	
	\subsubsection{Identity-mapping loss function}
The identity mapping loss function, shown in Eq. \eqref{eq:idm}, is used to enhance the generator's capacity to preserve the phonetic structure of the input speech. To calculate this loss, the source spectral feature is fed into the generator, where an attribute vector is assigned to indicate the source environment. Consequently, the generator's output should ideally match the input, with the loss calculated based on any discrepancies.
	\begin{equation}\label{eq:idm}
			\begin{aligned}
				\mathcal{L}_{idm}^{\mathcal{G}}={E}\{\| \mathcal{G}(\mathbf{Y}, \mathbf{t})-\mathbf{Y}\|_1\}.
			\end{aligned}
	\end{equation}

\subsection{InQSS}
Recently, several DL-based methods to predict speech quality and intelligibility have been proposed \cite{zhang2021end, spille2018predicting, manocha2021noresqa, hao2023neural}, showcasing high prediction performances owing to the robust modeling capabilities of DL-based models. InQSS is a multi-task speech assessment model designed to simultaneously predict human perception scores for speech quality and intelligibility. In particular, InQSS utilizes spectral and scatter coefficients to fully incorporate speech characteristics. As reported in \cite{chen2021inqss}, InQSS has been shown to outperform related methods by utilizing multiple acoustic features and a multitask learning criterion. Additionally, since most speech assessment datasets are in English, InQSS stands out as a valuable asset in the realm of speech assessment owing to its unique training on a Mandarin corpus. This corpus encompasses a diverse range of speech data, including clean, noisy, and enhanced speech processed through several SE methods. Each sample is accompanied by corresponding scores for speech quality and intelligibility.

The InQSS model utilized in this study is based on the CNN-bidirectional long short-term memory (BLSTM) architecture, which comprises two distinct stages. The first stage focuses on signal processing, extracting two streams of acoustic features from the input speech: spectral features and scattering coefficients \cite{mallat2012group}. The scattering coefficient, obtained through the application of the scattering transform to the speech waveform, is known for its robustness, time-shift invariant, and informational value. These acoustic features are then fed into CNN models for further processing. The outcomes of the two CNN models in the first stage are concatenated and utilized in the second stage, which employs the BLSTM model architecture to predict speech quality and intelligibility scores. Evaluation results in \cite{chen2021inqss} have demonstrated that the InQSS system offers more precise and dependable predictions for speech intelligibility and quality on Mandarin test sets compared with those of related assessment methods.

\begin{figure}[!tp]
    \centering
    \includegraphics[scale=\FigScal]{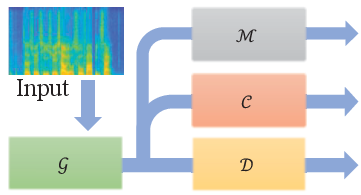}
    \caption{Block diagram of the proposed HL--StarGAN, comprising (1) generator $\mathcal{G}$, (2) discriminator $\mathcal{D}$, (3) classifier $\mathcal{C}$, and (4) metric predictor $\mathcal{M}$.}
    \label{fig:ccafv}
\end{figure}

\begin{figure*}[!tp]
    \centering
    \includegraphics[scale=\FigScal]{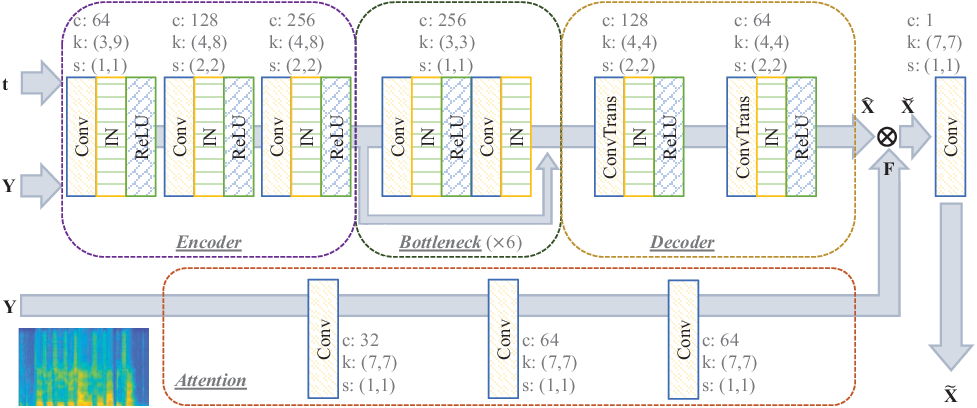}
    \caption{The detailed model structure of the generator (in HL--StarGAN) comprises encoder, bottleneck, decoder, and attention modules. The system input is the face-masked spectra, $\mathbf{Y}$, and the target attribute, $\mathbf{t}$, while the output is the processed spectra.}
    \label{fig:generator}
\end{figure*}

\section{Proposed Approach}\label{sec:ccafv}
The diagram of the proposed HL-StarGAN method, comprising four subsystems: (1) generator $\mathcal{G}$, (2) discriminator $\mathcal{D}$, (3) classifier $\mathcal{C}$, and (4) metric predictor $\mathcal{M}$ is shown in Fig. \ref{fig:ccafv}. The MaskQSS model is utilized to implement the $\mathcal{M}$ predictor. In this research, the log power spectrum (LPS) is utilized as the speech feature. Notably, during the inference phase, only the generator was employed to perform face-masked SE. Detailed explanations of the four subsystems are presented hereinafter.

\subsection{Generator}
The detailed model structure of the generator, comprising an encoder, bottleneck, decoder, and attention block is shown in Fig \ref{fig:generator}. Two-dimensional convolution (denoted as ``Conv''), transposed convolution (represented as ``ConvTrans''), and instance normalization (denoted as ``IN'') operations are integral components of this block. The parameters c, k, and s in the figure represent the number of channels, kernel size, and stride of the applied Conv (or ConvTrans) layer, respectively. LPS ($\mathbf{Y}$) and target attribute vector $\mathbf{t}$ serve as the inputs to the generator. The target attribute vector is a one-hot vector that indicates whether the output speech is clean (face-mask-free) or masked with a specific type of face mask.

As shown in Fig. \ref{fig:generator}, the generator generates speech spectra of the target domain through two paths. In the first path, the encoder processes the input spectral feature $\mathbf{Y}$ and attribute vector $\mathbf{t}$. The output features of the encoder pass through the bottleneck block to extract speech representations and generate the spectral feature $\hat{\mathbf{X}}$. The target domain in this study can be one of the three types of masks or a clean condition, resulting in a four-dimensional attribute vector. In the second path, the spectral feature $\mathbf{Y}$ is inputted into the attention block, generating a filter $\mathbf{F}$. Finally, $\hat{\mathbf{X}}$ is applied to filter $\mathbf{F}$, which can be expressed as follows:
 \begin{equation}\label{eq:attdec}
\breve{\mathbf{X}}=\hat{\mathbf{X}}\otimes\mathbf{F},
\end{equation}
where ``$\otimes$'' represents element-wise multiplication. Finally, generated output spectra are obtained by feeding $\breve{\mathbf{X}}$ through a convolution layer that undergoes post-processing to generate $\tilde{\mathbf{X}}$. Notably, during the testing phase, only the attribute vector $\mathbf{t}$ is specified as clean (face-mask-free) to guide the model in producing the face-mask-free spectral output.

\subsection{Classifier and discriminator}
\begin{figure}[!tp]
    \centering
    \includegraphics[width=\FIGWidtS]{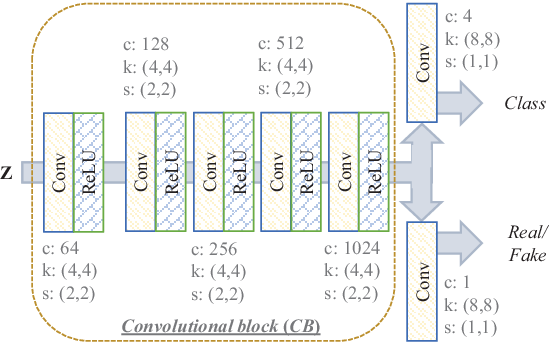}
    \caption{The classifier and discriminator model structures in HL--StarGAN are designed using the multi-task learning criterion.}
    \label{fig:clsdis}
\end{figure}
The system that integrates the classifier and discriminator of HL--StarGAN, implemented using a multitask criterion is shown in Fig \ref{fig:clsdis}. The system comprises a convolutional block (denoted as 'CB') and two distinct convolutional layers: one serving as a classifier and the other as a discriminator. The CB contains four hidden convolutional layers that extract latent features used for the types of face masks and determine the real or fake scores based on the classifier and discriminator. Real or fake identities are obtained as a sequence of one-dimensional frames and averaged to provide the overall result for each utterance. The generation of the classifier output $\mathbf{c}$ and discriminator output $d$ is expressed in Eq. \eqref{eq:clsdis}.
\begin{equation}\label{eq:clsdis}
    \begin{aligned}
       \mathbf{c}&=\mathrm{Conv}_1\{\mathrm{CB}\{\mathbf{Z}\}\},\\
        d&=\frac{1}{N}\sum_{T}\mathrm{Conv}_2\{\mathrm{CB}\{\mathbf{Z}\}\},
    \end{aligned}
\end{equation}
where $\mathrm{Conv}_1$ and $\mathrm{Conv}_2$ represent two distinct convolution layers, and $\mathbf{Z}$ can be clean speech or enhanced speech.

\subsection{MaskQSS}\label{sec:maskqss}
\begin{figure}[!tp]
    \centering
    \includegraphics[width=\FIGWidtS]{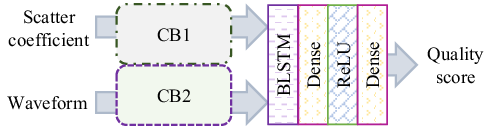}
    \caption{The MaskQSS model structure is employed for estimating the MOS score of an input speech that is clean or distorted by a face mask.}
    \label{fig:maskqss}
\end{figure}
In this section, we present the proposed MaskQSS model, developed to predict the quality scores of face-masked speech. The structure of MaskQSS, shown in Fig. \ref{fig:maskqss}, comprises two convolutional blocks (CB1 and CB2), a BLSTM layer, and a feed-forward (Dense) network. The CB modules comprise multiple convolutional layers, with the specific configurations listed in Table \ref{tab:maskqss}. Following the CB modules, the network architecture progresses with 512-node BLSTM layer, 128-node dense layer, and 1-node feed-forward layer, with a ``ReLU'' activation function applied to each hidden layer. The inputs to MaskQSS include scattering coefficients and speech waveforms. The CB1 module processes the scattering coefficients \cite{ghezaiel2021hybrid, moufidi2022wavelet, zeghidour2016deep, anden2014deep}, whereas the CB2 module processes the speech waveform to obtain latent representations. The outputs of these two modules are concatenated to create an 896-dimensional acoustic representation, which is then inputted into the subsequent BLSTM-Dense network. Finally, MaskQSS generates the MOS scores in the form of a 1D sequence. An averaging operation is performed on the output sequence to obtain a single speech quality score.

In this study, the MaskQSS model adopts a CNN-BLSTM architecture, akin to the InQSS and MOSNet frameworks. While MaskQSS utilizes scattering coefficients and raw waveforms as inputs, InQSS utilizes scattering coefficients and spectral features, and MOSNet relies solely on spectral features. In addition, MaskQSS outputs a predicted quality score particularly tailored for face-masked speech, whereas MOSNet evaluates the quality of synthesized speech of VC or TTS systems. On the other hand, InQSS predicts quality and intelligibility values for noisy speech, employing two BLSTMs to output scores, resulting in a larger model size compared with that of MaskQSS.

\begin{table}[!bp]
\caption{The (channel number, kernel size, stride) setup of nine convolutional hidden layers in ``CB1'' and ``CB2.''}\label{tab:maskqss}
\centering
\begin{tabularx}{\TabWidt}{>{\centering}m{\TabCWidF}|>{\centering}m{\TabCWidF}>{\centering\arraybackslash}X}
\toprule \hline
\textbf{LayerID}&\textbf{CB1}&\textbf{CB2}\\
\hline
(1)&\multicolumn{2}{c}{16, 3, (1,1)}\\
(2)&\multicolumn{2}{c}{16, 3, (1,1)}\\
(3)&\multicolumn{2}{c}{16, 3, (1,3)}\\
(4)&\multicolumn{2}{c}{32, 3, (1,1)}\\
(5)&\multicolumn{2}{c}{32, 3, (1,3)}\\
(6)&\multicolumn{2}{c}{64, 3, (1,1)}\\
(7)&\multicolumn{2}{c}{64, 3, (1,3)}\\
(8)&\multicolumn{2}{c}{128, 3, (1,1)}\\
(9)&\multicolumn{2}{c}{128, 3, (1,3)}\\
\hline \bottomrule
\end{tabularx}
\end{table}

\subsection{Training process}\label{sec:trainingprocess}
\begin{figure}[!tp]
    \centering
    \includegraphics[width=\FIGWidtT]{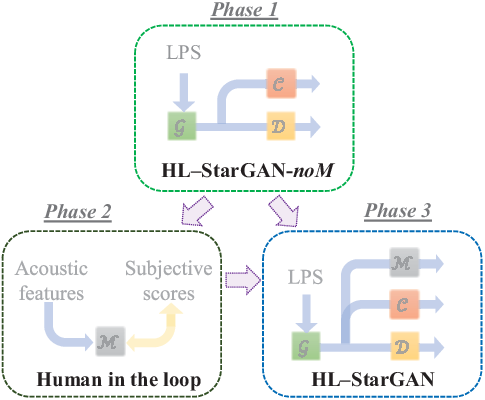}
    \caption{Learning HL--StarGAN involves three phases: (a) HL--StarGAN-\textit{noM}, (b) human-in-the-loop, and (c) HL--StarGAN.}\label{fig:traproc}
\end{figure}
The training process of the proposed HL--StarGAN model involves three phases, as shown in Fig. \ref{fig:traproc}. In Phase 1, HL--StarGAN is trained without the $\mathcal{M}$ module, referred to as ``HL--StarGAN-\textit{noM}.'' The model is learned using the functions in Eq. \eqref{eq:all_loss} to calculate the loss during this phase, where the function $\mathcal{L}_{noM}$ is defined in Eq. \eqref{eq:gcd_loss}.
\begin{equation}\label{eq:all_loss}    \mathcal{L}_{noM}=\mathcal{L}^{\mathcal{C},\mathcal{D}}_{noM}+\mathcal{L}^{\mathcal{G}}_{noM},
\end{equation}
where
\begin{equation}\label{eq:gcd_loss}
\begin{aligned}    \mathcal{L}^{\mathcal{C},\mathcal{D}}_{noM}&=\mathcal{L}_{adv}^\mathcal{D}+\lambda_1\mathcal{L}_{cls}^\mathcal{C},\\
    \mathcal{L}^{\mathcal{G}}_{noM}&=\mathcal{L}_{adv}^\mathcal{G}+\lambda_2\mathcal{L}_{cyc}^\mathcal{G}+\lambda_3\mathcal{L}_{cls}^\mathcal{G}+\lambda_4\mathcal{L}_{idm}^\mathcal{G}.
    \end{aligned}
\end{equation}
In Eq. \eqref{eq:gcd_loss}, the hyperparameters $\lambda_1$, $\lambda_3$, and $\lambda_4$ were set to 2, while $\lambda_2$ was set to 3 in this study. Subsequently, the loss function expressed in Eq. \eqref{eq:all_loss} was optimized to iteratively update the discriminator, classifier, and generator.

In Phase 2, referred to as the ``human-in-the-loop" phase, training was conducted utilizing the metric prediction module $\mathcal{M}$ using the MaskQSS, as discussed in Section \ref{sec:maskqss}. During this phase, MaskQSS was trained using paired utterance-MOS training data, which included face-masked speech processed by generator $\mathcal{G}$ in HL--StarGAN-\textit{noM} and clean speech. Ground truth MOS scores were obtained through subjective listening tests conducted by recruited listeners who rated the speech generated by the HL--StarGAN-\textit{noM} generator. During the training of MaskQSS, the training utterances were passed through MaskQSS to generate MOS predictions. The L1 distance was utilized to measure the difference between the predictions and ground truth MOS. The L1 distance was also employed as the loss function to optimize the MaskQSS model, expressed as follows:
\begin{equation}\label{eq:mos_loss}
    \mathcal{L}_{MOS}^{\mathcal{M}}=E\{\|\mathcal{M}\{\tilde{\mathbf{x}}\}-MOS\|\},
\end{equation}
where $\tilde{\mathbf{x}}$ and $\mathcal{M}\{\cdot\}$ denote the speech and mapping function of MaskQSS, respectively. Notably, $\mathcal{G}$ generates the target-domain spectral features (LPS), which were then processed by an inverse Fourier transform to convert them back to a time-domain signal $\tilde{\mathbf{x}}$ for MaskQSS. Finally, the trained MaskQSS was employed as a human-centric quality predictor and utilized in the third phase to train the HL--StarGAN.

In Phase 3, HL--StarGAN-\textit{noM} was integrated with MaskQSS to form HL--StarGAN. This integration was followed by further training using the loss function outlined in Eq. \eqref{eq:hstargan}.
\begin{equation}\label{eq:hstargan}
    \mathcal{L}=\mathcal{L}_{noM}+\mathcal{L}_{5.0}^\mathcal{M}.
\end{equation}
To define the loss function $\mathcal{L}_{5.0}^\mathcal{M}$ in this equation, we assigned the $MOS$ term of $\mathcal{L}_{MOS}^{\mathcal{M}}$ in Eq. \eqref{eq:mos_loss} with the highest MOS score of 5.0. The discriminator, classifier, generator, and MaskQSS were iteratively updated within each epoch to effectively train HL--StarGAN for the face-masked SE task. Notably, at this phase and during inference, the clean (face-mask-free) condition was designated as the target attribute $\mathbf{t}$, as shown in Fig. \ref{fig:generator}, guiding the generator to generate face-mask-free speech. 

After Phase 3, the trained HL--StarGAN model goes through Phase 2 again for the next round of training. After several iterations of Phase 2 and Phase 3, we obtain the final HL--StarGAN model.

\begin{table*}[!t]
\centering
\caption{The detailed configuration of the face-masked speech database}\label{tab:database}
\begin{tabularx}{\textwidth}{r|>{\raggedright\arraybackslash}m{16cm}}
\toprule \hline
\textit{\textbf{Attributes}}&{\hspace{7cm}\textit{\textbf{Descriptions}}}\\
\hline
\hline
\textbf{Speakers}&\begin{itemize}
    \item \textit{Number}: 15 female and 19 male
    \item \textit{Age}: $20\sim 30$ years old
    \end{itemize}\\
\hline
    \textbf{Conditions}&\begin{itemize}
        \item \textit{Face mask types}: ``N95,'' ``cotton,'' ``plastic shield''
        \item Clean
    \end{itemize}\\
    \hline
    \textbf{Recordings}&\begin{itemize}
        \item \textit{Equipment}: a microphone in the laptop
        \item \textit{Sampling rate}: 48000 Hz
        \item \textit{Size}: 43,520 utterances ($\sim 42$ hrs.)
    \end{itemize}\\
    \hline
        \textbf{Environments}&\begin{itemize}
            \item $2 \times2 \times 2$ (meter) semi-anechoic room
        \item \textit{Distance} (\textit{from microphone to a speaker}): $(Horizontal, vertical)=(30,40)$ (centimeter) 
        \end{itemize}\\
    \hline
            \textbf{Corpus}&\begin{itemize}
                \item TMHINT
                \item \textit{320 sentences}: 290 training, 10 validation and 20 testing sentences
                \item \textit{Each sentence}: 10 Chinese characters
            \end{itemize}\\
    \hline\bottomrule
\end{tabularx}
\end{table*}

\section{Experiments and analyses}\label{sec:ea}
\subsection{Experimental setup}
\begin{figure}[!t]
    \centering
    \includegraphics[width=\FIGWidtF]{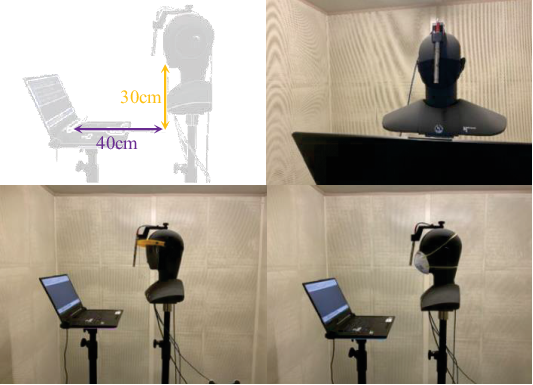}
    \caption{Physical recording environment and configurations}
    \label{fig:phyrec}
\end{figure}

Our experiments were conducted using the FMVD database, a collection of face-masked speech recordings. The sentences in this database were selected from the Taiwan Mandarin noise listening test (TMHINT) corpus \cite{huang2005development}. The training set consisted of 290 sentences, the validation set had 10 sentences, and the test set included 20 sentences. Each sentence contained 10 Chinese characters. The FMVD database featured recordings from 15 female and 19 male speakers wearing three types of masks, namely ``N95,'' ``cotton,'' and ``plastic shield.'' Participants wore masks during the recording process and recited sentences to generate utterances. Clean waveforms were recorded for all 34 speakers, resulting in approximately 42 hours of audio across 43,520 utterances in the FMVD database. Recordings were captured using a laptop microphone in a semi-anechoic chamber at a sampling rate of 48 \textit{k}Hz, which was later downsampled to 16 \textit{k}Hz. More detailed information on the FMVD database configuration is found in Table \ref{tab:database}. The physical recording setup is shown in Fig. \ref{fig:phyrec}. In this study, a subset of 25 speakers and their associated utterances were randomly selected for the training and validation datasets. The test set included utterances from two additional male and female speakers. In total, 29,000, 100, and 240 utterances were utilized for training, validation, and evaluation of the proposed HL--StarGAN system, respectively.

Except for speech evaluation models that utilized waveform inputs, the short-time Fourier transform was employed to extract LPS from speech signals. The LPS was obtained using a window size of 512 points and a hop length of 80 points, resulting in 257-dimensional features in the frequency domain. The HL--StarGAN-\textit{noM}, MaskQSS, and HL--StarGAN models were trained in three phases, as previously mentioned in Section \ref{sec:trainingprocess}, using a GeForce RTX 1080 GPU. Throughout the entire training process, the Adam optimizer with $\beta_1=0.5$ and $\beta_2=0.999$ was utilized for model training. The learning rate was set to 0.0001, and the batch size was 2. The models in the three training phases were subjected to 200,000 iterations. We employed DNSMOS \cite{DNSMOS} and subjective tests to assess the performance of the proposed system.

\subsection{Subjective test}
\subsubsection{Listening test on speech quality}
To conduct the subjective listening tests for the second phase in Section \ref{sec:trainingprocess}, the HL--StarGAN system was first trained without MaskQSS (HL--StarGAN-\textit{noM}) on the FMVD database. Subsequently, we collected 120 face-masked utterances by randomly selecting 40 utterances in the FMVD from each face-masked scenario. These face-masked utterances were then processed using HL--StarGAN-\textit{noM} to obtain enhanced speech. Fifteen participants with normal hearing were recruited for subjective testing. In a quiet room, they were requested to rate the quality of both face-masked and HL--StarGAN-enhanced utterances on a scale ranging from 1 to 5, with higher scores indicating better quality. The quality values obtained were averaged, resulting in 240 scores (120 face-masked speech and 120 HL--StarGAN-\textit{noM}-enhanced utterances). The quality scores for the HL--StarGAN-\textit{noM}-enhanced utterances were labeled as ``HL--StarGAN-\textit{noM}'' whereas those for the face-masked utterances were labeled as ``Masked speech.'' The results for ``Masked speech'' and ``HL--StarGAN-\textit{noM}'' are listed in Table \ref{tab:mosscore}. Additionally, participants were tasked with rating the quality of 40 unprocessed clean speech samples for comparison, resulting in an average score of 3.93.

\begin{table}[!b]
    \caption{Subjective test scores for ``Masked speech'' and ``HL--StarGAN-\textit{noM}'' in each of the three distinct face-masked scenarios.}\label{tab:mosscore}
  \begin{center}
    \begin{tabularx}{\TabWidt}{>{\raggedleft}m{\TabCWidS}|>{\centering}m{\TabCWidT}>{\centering}m{\TabCWidT}>{\centering\arraybackslash}X}
      \toprule \hline
      & \textbf{Cotton} & \textbf{N95} & \textbf{Plastic}\\
      \hline
        \textbf{Masked speech} & 3.40 & 3.61 & 3.41 \\ 
        \textbf{HL--StarGAN-\textit{noM}} & \textbf{3.81} & \textbf{3.86} & \textbf{3.76} \\
      \hline \bottomrule
    \end{tabularx}
  \end{center}
\end{table}

The results presented in Table \ref{tab:mosscore} reveal that the scores for ``Masked speech'' are considerably lower compared with those of the clean condition, indicating that masks can negatively influence speech quality. Next, the application of the HL--StarGAN enhancement algorithm significantly enhanced the score of the ``Masked Speech,'' underscoring the effectiveness of the proposed HL--StarGAN method in improving the auditory perception of human voices.

\subsubsection{Listening test on paired comparison}
\begin{figure}[!tp]
    \centering
    \includegraphics[width=\FIGWidtT]{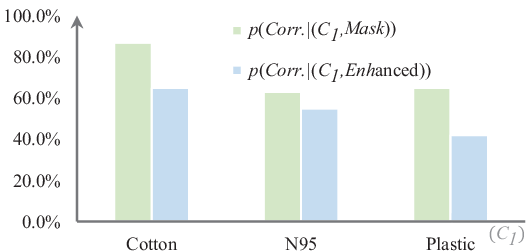}
    \caption{Accuracy in subjectively perceiving face-masked voices. A lower accuracy suggests difficulty in differentiating between mask-distorted or HL--StarGAN-enhanced speech and clean reference speech.}
    \label{fig:abtest}
\end{figure}

In addition to the speech quality test, participants were recruited to partake in listening tests aimed at distinguishing between voices spoken without face masks, those spoken with masks, and those generated by HL--StarGAN. For this evaluation, 40 pairs of voice waveforms were generated, each containing either \{clean speech; face-masked speech\} or \{clean speech; HL--StarGAN-enhanced speech\}. To ensure unbiased test results, the speech content and speakers differed for both voices in each pair. The HL--StarGAN system utilized to generate the enhanced speech was consistent with the one employed in the previously mentioned subjective test. Fifteen participants, unaware of the condition of each testing pair, participated in the subjective test. During the experiment, participants were presented with a pair of voices and tasked with determining whether the provided utterances were distorted by a face mask. If they answered positively, they were then required to specify the affected voice(s)--first, second, or both. Finally, based on the specifications of all 80 utterances, the accuracy ratio was calculated using Eq. \eqref{eq:acc}, and the results are shown in Fig. \ref{fig:abtest}.
\begin{equation}\label{eq:acc}
    p(Corr.|(C_1,C_2))=\frac{\# Correct\;answer}{\#(C_1\;\&\;C_2)\times (15 subjects)},
\end{equation}
where $C_1\in\{\mbox{``N95''},\mbox{ ``cotton''}, \mbox{``plastic shield''}\}$ and $C_2\in\{Mask, Enhanced\}$ indicated the voice condition.

As shown in Fig. \ref{fig:abtest}, listeners could easily distinguish between cotton-distorted and clean utterances compared with other speech pairs. This finding suggests that the fiber structure of cotton masks significantly influences the perception of human speech. Further, we observed that the accuracy of identifying $p(Corr.|(C1, Enhanced))$ is lower than that of $p(Corr.|(C1, Mask))$, indicating the difficulty in differentiating between processed face-masked and clean speech. These results validate the effectiveness of the HL--StarGAN algorithm in enhancing face-masked speech to produce speech that cannot be easily distinguished from face-mask-free speech.

\subsection{MaskQSS}
\begin{figure}[!tp]
    \centering
    \includegraphics[width=\FIGWidtT]{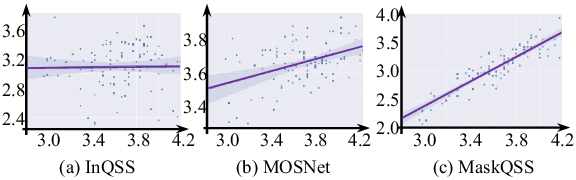}
    \caption{Prediction of human perceptual scores using (a) InQSS, (c) MOSNet, and (d) MaskQSS models. The $x$-axis represents the predicted results, whereas the $y$-axis represents the human perception scores. The $y$-axis range differs among the three subfigures.}\label{fig:corr}
\end{figure}

Next, we evaluated the effectiveness of the MaskQSS module in predicting human perceptions of face-masked voices. The evaluation utilizes the ``sFMVD'' database, which is a subset of FMVD containing spoken sentences and corresponding MOS scores obtained from the previous subjective test. Within the sFMVD dataset, 160 face-masked and clean utterances, along with their associated averaged MOS scores, were selected for evaluation. To train and test the MaskQSS module, we adopted the second phase of the training process outlined in Section \ref{sec:trainingprocess} and employed the five-fold cross-validation technique. Furthermore, the performances of two additional models, MOSNet and InQSS, were assessed using the same test set. Notably, InQSS was trained on the TMHINT-QI database \cite{chen2021inqss}, whereas MOSNet was pretrained and then fine-tuned on the sFMVD face-masked voices. The evaluation results are shown in Fig. \ref{fig:corr}.

As shown in Fig. \ref{fig:corr}, the scatter plots, which serve as the qualitative analysis, reveal positive correlations between the MOS (human subjective listening test) and MOSNet, and MOS and MaskQSS predictions. This observation indicates that MOSNet and the proposed MaskQSS model can effectively predict quality scores that are in line with human perception. Upon closer examination of Figs. \ref{fig:corr} (a) with (b) and (c), both MOSNet and the proposed MaskQSS system generated a wide score range, with MaskQSS exhibiting the widest range of prediction scores, indicating that the MaskQSS approach is proficient in generating highly discriminative quality scores, making it valuable for evaluating speech in the context of face masks.

\begin{table}[!b]
    {
    \caption{PCC and SRCC were calculated to determine the relationship between human perception scores and model predictions.}\label{tab:corr}
	\centering
\begin{tabularx}{\TabWidt}{>{\raggedleft}m{\TabCWidT}|>{\centering}m{\TabCWidu}>{\centering\arraybackslash}X}
      \toprule \hline
      \textbf{Model} & \textbf{PCC} & \textbf{SRCC} \\
      \hline
      \textbf{InQSS} & 0.03 & 0.01\\
      \textbf{MOSNet} & 0.45 & 0.34\\
      \textbf{MaskQSS} & \textbf{0.91} & \textbf{0.88} \\
      \hline \bottomrule
    \end{tabularx}
}
\end{table}

Additionally, we analyzed several correlation scores, such as the PCC and the Spearman rank correlation coefficient (SRCC) to quantitatively evaluate the correlation between human perception scores and model predictions. The resulting correlations are listed in Table \ref{tab:corr}, according to which the InQSS model yielded the lowest correlations whereas the MaskQSS-generated scores showed a strong correlation with human perceptions. These outcomes underscore the effectiveness of the MaskQSS model in predicting the quality scores of face-masked voices.

\begin{figure}[!tp]
    \centering
    \includegraphics[width=\FIGWidto]{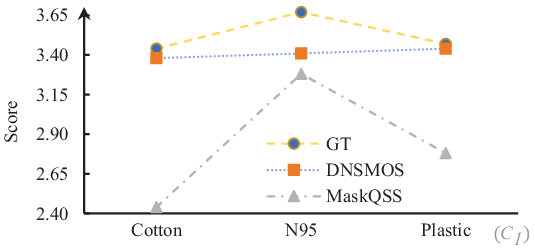}
    \caption{Evaluation of the speech quality of the recordings in the FMVD database subjected to mask corruption using the MaskQSS and DNSMOS metrics. In addition, the MOS scores of the subjective test are listed for comparison (denoted as ``GT'').}\label{fig:dnsqsscmp}
\end{figure}

Next, we utilized the trained MaskQSS to assess the audio quality of the test recordings in the FMVD dataset and compared its performance with the average DNSMOS scores for face-masked utterances. The results are shown in Fig. \ref{fig:dnsqsscmp}, with the $C_1$ axis, representing the three mask scenarios. The abbreviations ``MaskQSS'' and ``DNSMOS'' denote the average scores obtained from the MaskQSS and DNSMOS functions, respectively. Additionally, the subjective MOS results from the tests (indicated as ``GT'') were included in this figure for comparison. When evaluating the scores along the $x$-axis, MaskQSS demonstrates a wider range of dynamic values for quality assessment compared with DNSMOS. Furthermore, a consistent score trajectory pattern was observed when comparing MaskQSS and GT along the $x$-axis. These observations suggest that the mask-specific MaskQSS provides more discriminative quality scores compared with the DNSMOS metric derived in noisy environments, making it a more suitable tool for evaluating speech quality impacted by face masks.

\subsection{HL--StarGAN performance evaluation}
This section presents the evaluation results of the proposed systems. First, we compared the performances of the conventional CycleGAN and HL--StarGAN methods using the DNSMOS metric, and the results are listed in Table \ref{tab:ccafvunsup}. The average scores computed for face-masked voices processed by HL--StarGAN are denoted as ``HL--StarGAN," whereas those for CycleGAN-enhanced voices are denoted ``CycleGAN." Additionally, the table includes DNSMOS values for the unprocessed voices, referred to as ``Masked Speech." Notably, the implementation of the unsupervised conditional CycleGAN is an SE approach based on the methodology outlined in \cite{ting2022speech}.

\begin{table}[!bp]

    \centering
    \caption{Averaged DNSMOS scores of CycleGAN, HL--StarGAN, and Masked speech under all testing scenarios.}\label{tab:ccafvunsup}
    \begin{tabularx}{\TabWidt}{>{\raggedleft}m{\TabCWidT}|>{\centering}m{\TabCWidi}>{\centering}m{\TabCWidi}>{\centering}m{\TabCWidi}>{\centering\arraybackslash}X}
       \toprule\hline
        & \textbf{Cotton} &  \textbf{N95}  & \textbf{Plastic}& \textbf{Avg.}\\
        \hline
        \textbf{Masked speech}&3.38&3.41&3.44&3.41\\  
        \textbf{CycleGAN}&3.46&3.47&3.47&3.47\\ 
        \textbf{HL--StarGAN}&\textbf{3.82}&\textbf{3.82}&\textbf{3.66}&\textbf{3.77}\\
        \hline\bottomrule
    \end{tabularx}
\end{table}

The comparison between HL–StarGAN and CycleGAN with the baseline masked speech in Table \ref{tab:ccafvunsup} indicates the effectiveness of utilizing these unsupervised learning methods to enhance face-masked speech across all scenarios. Moreover, the superior performance of HL--StarGAN suggests that incorporating a classifier and ``human-in-the-loop'' metric prediction module can further enhance system performance compared with that of the generator-discriminator (i.e. CycleGAN) architecture.

To validate the effectiveness of the speech assessment model, we trained an additional HL--StarGAN model, referred to as ``HL--StarGAN(I),'' for comparison. In {HL--StarGAN(I)}, we replaced MaskQSS in the HL--StarGAN model with InQSS to perform the $\mathcal{M}$ function, as shown in Fig. \ref{fig:ccafv}. Subsequently, both the HL--StarGAN and HL--StarGAN(I) models were utilized to enhance the face-masked utterances in the FMVD testing set, generating corresponding outputs. These outputs were then evaluated using DNSMOS to calculate the quality scores. Furthermore, the proposed MaskQSS was applied to compute the perception scores for enhanced speech. The metric scores for the HL--StarGAN and HL--StarGAN(I) methods were averaged and labeled the resulting values as ``HL--StarGAN'' and ``HL--StarGAN(I),'' respectively. These results are listed in Table \ref{tab:ccafvim}. To establish a baseline for comparison, DNSMOS and MaskQSS scores were also computed for the unprocessed face-masked voices, referred to as ``Masked speech.''

\begin{table}[!bp]
    \centering
    \caption{Averaged DNSMOS and MaskQSS scores of {HL--StarGAN(I), HL--StarGAN}, and Masked speech in the FMVD testing set.}\label{tab:ccafvim}
    \begin{tabularx}{\TabWidt}{>{\raggedleft}m{\TabCWidT}|>{\centering}m{\TabCWidu}>{\centering\arraybackslash}X}
       \toprule\hline
        & \textbf{DNSMOS} &  \textbf{MaskQSS}\\
        \hline
        \textbf{Masked speech}&3.41&3.23\\  
        \textbf{{HL--StarGAN(I)}}&3.62&3.29\\  
        \textbf{HL--StarGAN}&\textbf{3.76}&\textbf{3.47}\\  
        \hline\bottomrule
    \end{tabularx}
\end{table}

As shown in Table \ref{tab:ccafvim}, HL--StarGAN and HL--StarGAN(I) demonstrated higher evaluation scores compared with the masked speech, with HL--StarGAN demonstrating superior quality performance. Therefore, combining a more accurate metric-prediction model with an enhancement system can improve face-masked SE performance. 

\begin{table}[!bp]
    \centering
    \caption{Averaged MaskQSS scores of HL--StarGAN(I), HL--StarGAN, and Mask under three specific scenarios.}\label{tab:ccafvimdetail}
    \begin{tabularx}{\TabWidt}{>{\raggedleft}m{\TabCWidT}|>{\centering}m{\TabCWidTa}>{\centering}m{\TabCWidTa}>{\centering\arraybackslash}X}
       \toprule\hline
        & \textbf{Cotton} &  \textbf{N95}  & \textbf{Plastic}\\
        \hline
        \textbf{Masked speech}&2.44&3.28&2.78\\  
        \textbf{{HL--StarGAN(I)}}&3.45&3.32&3.10\\
        \textbf{HL--StarGAN}&\textbf{3.62}&\textbf{3.43}&\textbf{3.36}\\
        \hline\bottomrule
    \end{tabularx}
\end{table}

The detailed performances of Masked speech, HL--StarGAN(I), and HL--StarGAN under the three distinct face-masked scenarios are shown in Table \ref{tab:ccafvimdetail}, with the quality scores represented based on the MaskQSS metrics. Across all conditions, HL--StarGAN consistently achieved the best performance, indicating the effectiveness of the proposed HL--StarGAN method in enhancing face-masked speech.

\begin{table}[!bp]
    \centering
    \caption{Ablation study conducted on HL–StarGAN to showcase the effectiveness and efficiency of the implemented modules. The left three columns indicate the DNSMOS scores; the right two columns denote the numbers of parameters and FLOPs.}\label{tab:label}
    \begin{tabularx}{\TabWidt}{>{\raggedleft}m{\TabCWidxa}|>{\centering}m{\TabCWidx}>{\centering}m{\TabCWidx}>{\centering}m{\TabCWidx}|>{\centering}m{\TabCWidxb}>{\centering\arraybackslash}X}
        \toprule\hline
        & \textbf{Cotton} &  \textbf{N95}  & \textbf{Plastic} &\textit{\textbf{\#Par.}}&\textit{\textbf{FLOPs}}\\
        \hline
        \textbf{HL--StarGAN}&\textbf{3.82}&\textbf{3.82}&\textbf{3.66}&$9.37\times 10^6$&$58.94\times 10^9$\\  
        \hline
        -\textbf{\textit{noM}}&3.60&3.66&3.63&$9.37\times 10^6$&$58.94\times 10^9$\\
        -\textbf{\textit{noMA}}&3.31&3.35&3.26&$9.06\times 10^6$&$43.44\times 10^9$\\
        \hline\bottomrule
    \end{tabularx}
\end{table}

To evaluate the effectiveness of the attention and MaskQSS modules in HL--StarGAN, we conducted an ablation study in which HL--StarGAN was compared with two modified versions: HL--StarGAN-\textit{noM} and HL--StarGAN-\textit{noMA}. HL--StarGAN-\textit{noM} excluded the $\mathcal{M}$ function, whereas HL--StarGAN-\textit{noMA} excluded both the attention module in the generator and the $\mathcal{M}$ function from HL--StarGAN, essentially representing the StarGAN baseline system. Test utterances from FMVD were processed using all three HL--StarGAN systems and the average DNSMOS score was calculated for each face-masked scenario. The results are listed in Table \ref{tab:label}, where the parameter values (\textit{\#Par.}) and the number of floating-point operations (\textit{FLOPs}) computed per sample in the generator to compare computational costs. Both HL--StarGAN-\textit{noM} and HL--StarGAN-\textit{noMA} demonstrated lower speech quality compared with HL--StarGAN across all face-masked scenarios. The introduction of the attention mechanism resulted in a 0.3 improvement in quality compared with the HL--StarGAN-\textit{noMA} baseline, with further enhancements observed with the inclusion of the MaskQSS module. These findings underscored the significance of integrating the attention and $\mathcal{M}$ functions to enhance the speech quality of HL--StarGAN during inference. In terms of computational efficiency, all generators had a similar order of magnitude of parameters, whereas the generator of HL--StarGAN-\textit{noMA} exhibited slightly lower FLOP values compared with its other two counterparts.

\begin{figure}[!t]
    \centering
    \includegraphics[width=\FIGWidtT]{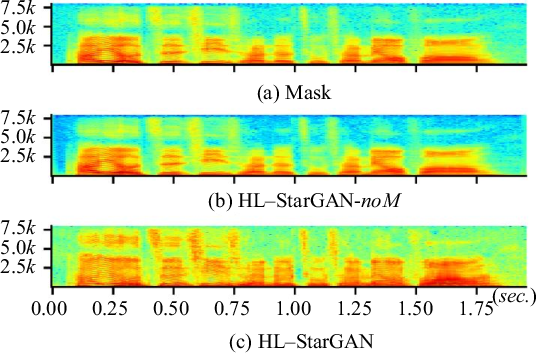}
    \caption{(a) Face-masked spectrogram, (b) HL--StarGAN-\textit{noM}- and (c) HL--StarGAN-processed face-masked speech. The unit of the $x$-axis is ``second,'' whereas that of the $y$-axis is ``Hz.''}
    \label{fig:spec}
\end{figure}

A comparison of the spectrograms of the original face-masked speech and processed speech, with the time and frequency displayed on the $x$- and $y$-axes, respectively, is shown in Fig. \ref{fig:spec}. A speech sample labeled as ``Masked speech'' was selected from the FMVD testing set and processed using two enhancement systems: HL--StarGAN and HL--StarGAN-\textit{noM}. The resulting output speeches were labeled as ``HL--StarGAN'' and ``HL--StarGAN-\textit{noM},'' respectively. As shown in Fig. \ref{fig:spec} (a), the energy of the Masked spectrum was primarily concentrated in the low-frequency range, resulting in a lack of clarity in the high-frequency region. Conversely, Fig. \ref{fig:spec} (b) shows that HL--StarGAN-\textit{noM} effectively reduced the background noise. On the other hand, as shown in Fig. \ref{fig:spec} (c), HL--StarGAN effectively enhanced the high-frequency components of the face-masked speech, resulting in more detailed sound structures. These findings highlighted the effectiveness of the HL--StarGAN-\textit{noM} technique in mitigating background noise, whereas HL--StarGAN, which utilized the MaskQSS module, primarily enhanced the high-frequency structure of face-masked speech.

\section{Conclusion}\label{sec:summary}
This study presented a novel method called the human-in-the-loop StarGAN (HL--StarGAN) face-masked SE system, aimed at enhancing the quality of speech recordings while wearing a mask. Furthermore, we developed the MaskQSS system to serve as the ``human-in-the-loop'' module for evaluating the quality score of face-masked speech. The HL--StarGAN system comprised generators, discriminators, classifiers, and a MaskQSS predictor. To train and test the proposed face-masked quality assessment and SE systems, we created a face-masked speech database called ``FMVD'' featuring recordings from 34 speakers in diverse clean and face-masked scenarios. The quality assessment of the HL--StarGAN system involved both subjective and objective tests. Subjective tests revealed a strong correlation between the human perception of speech quality and MaskQSS scores, indicating the effectiveness of MaskQSS in predicting MOS values for face-masked speech. The objective results demonstrated that when combined with MaskQSS, the HL--StarGAN system outperformed the baseline StarGAN approach in enhancing speech quality. Furthermore, our ablation study revealed that the attention module could boost the proposed system to more effectively enhance face-masked sounds. In future studies, we plan to explore various optimizations of the enhancement model, such as the use of alternative neural network architectures, different normalization schemes, and a comparison of attention mechanisms to determine the optimal design for performance and efficiency in online inferences.

\bibliographystyle{ieeetr}
\bibliography{my_bib}

\begin{IEEEbiography}
[{\includegraphics[width=1in,clip,keepaspectratio]{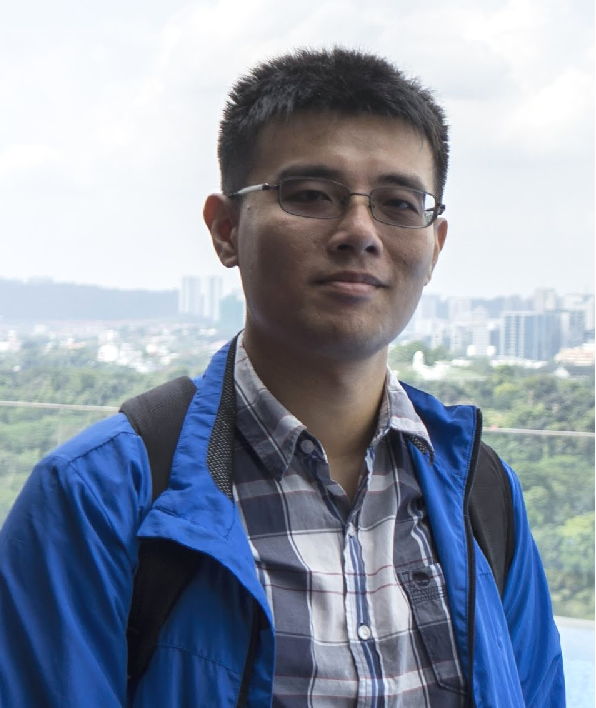}}]
{Syu-Siang~Wang} completed his Bachelor's (2008) and Master's (2010) degrees in Electrical Engineering from National Changhua Normal University in Changhua, Taiwan, and National Jinan University in Nantou, Taiwan, respectively. He obtained his Ph.D. from the Graduate Institute of Communication Engineering at National Taiwan University in 2018. During his Ph.D. studies, he served as a research assistant at the Information Technology Innovation Research Center of Academia Sinica from 2012 to 2018, focusing on robust speech feature extraction and enhancement. After completing his Ph.D., he conducted postdoctoral research at the NTU Joint Research Center for AI Technology and All Vista Healthcare, where he worked on hearing aid signal processing technology. Currently, he holds the position of assistant professor in the Department of Electrical Engineering at Yuan Ze University in Taoyuan, Taiwan. His research interests encompass speech signal processing and biological signal processing.
\end{IEEEbiography}
\begin{IEEEbiography}
[{\includegraphics[width=1in,clip,keepaspectratio]{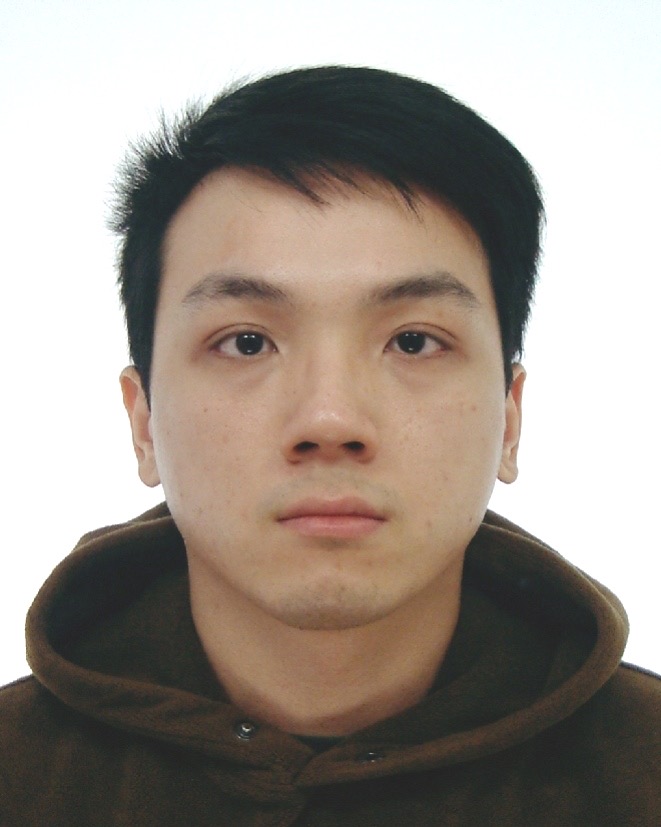}}]
{Jia-Yang Chen} received his B.S. degree in Electrical Engineering from Chang Gung University, in 2020, and master degree in Yuan Ze University, in 2022, respectively. He is currently working at MediaTek.inc. During the master, his research interests include speech signal processing, and machine learning.
\end{IEEEbiography}
\begin{IEEEbiography}
[{\includegraphics[width=0.93in,clip,keepaspectratio]{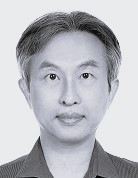}}]
{Bo-Ren Bai} received his B.S. and Ph.D. degrees in electrical engineering from National Taiwan University, in 1992 and 1998, respectively. After graduating, he began working in the speech processing software and IC design industries. He is currently the vice president of R\&D at Fortemedia. His research interests include speech signal processing, microphone arrays, human-machine voice interfaces, and machine learning. 
\end{IEEEbiography}
\begin{IEEEbiography}
[{\includegraphics[width=0.93in,clip,keepaspectratio]{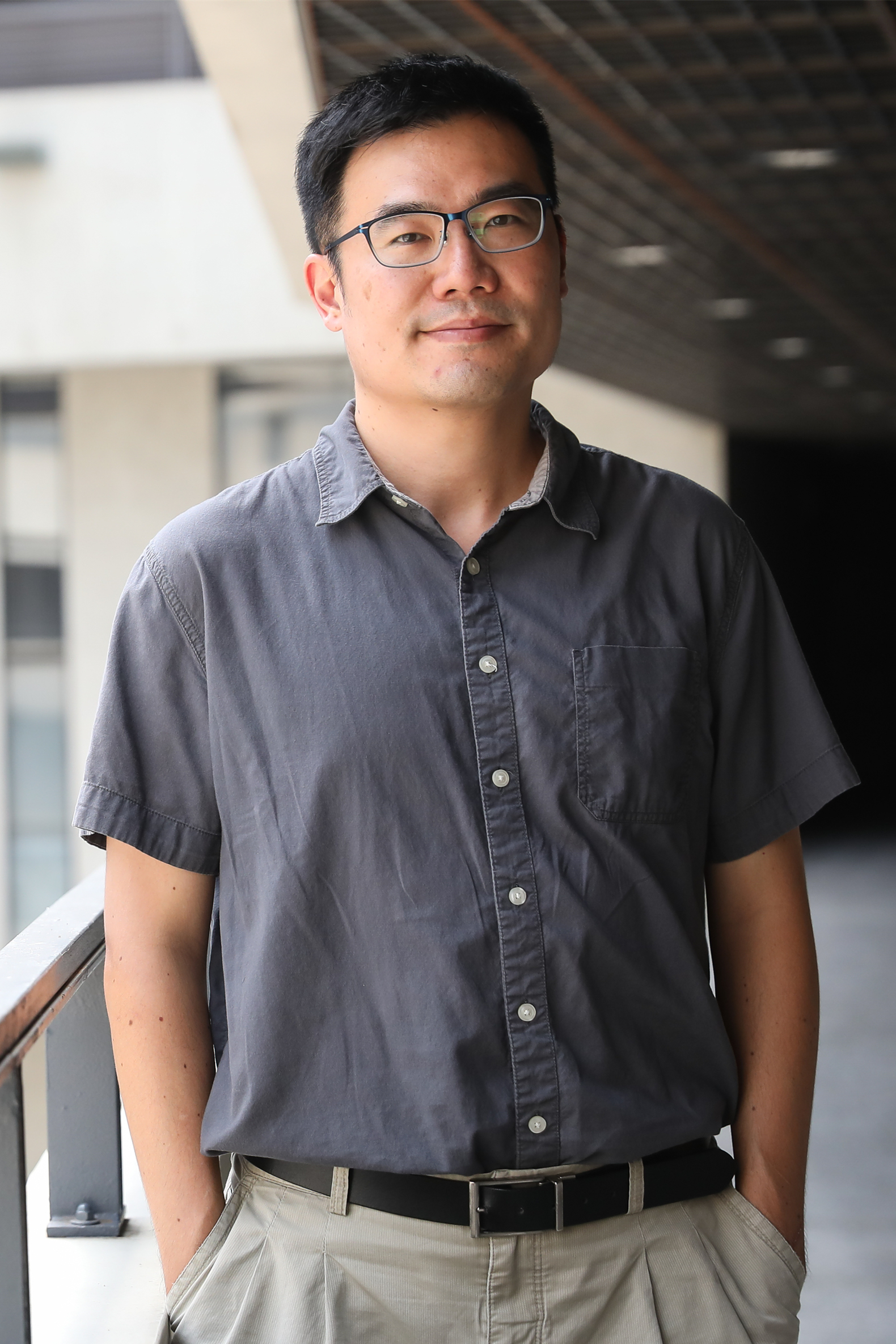}}]
{Shih-Hau Fang} (M’07-SM’13) is currently a Professor in the Department of Electrical Engineering, National Taiwan Normal University (NTNU) and Yuan Ze University (YZU). Prof. Fang’s research interests include mm-wave radar and acoustic signal sensing. He has authored or co-authored hundreds technical journal and conference papers in these fields. He is a fellow of IET and a senior member of IEEE. Prof. Fang has received several awards for his research work, including the Project for Excellent Junior Research Investigators (MOST, 2013), Outstanding Young Electrical Engineer Award (Chinese Institute of Electrical Engineering, 2017), Outstanding Research Award (YZU, 2018), Best Synergy Award (Far Eastern Group, 2018), Future Technology Award (MOST, 2019), National Innovation Award (RBMP, 2019), Y.Z Outstanding Professor (Y.Z. Hsu Science and Technology Memorial Foundation, 2019), Outstanding Electrical Professor (Chinese Institute of Electrical Engineering, 2021), and Y.Z Chair Professor (Y.Z. Hsu Science and Technology Memorial Foundation, 2021)
\end{IEEEbiography}
\begin{IEEEbiography}
[{\includegraphics[width=0.93in,clip,keepaspectratio]{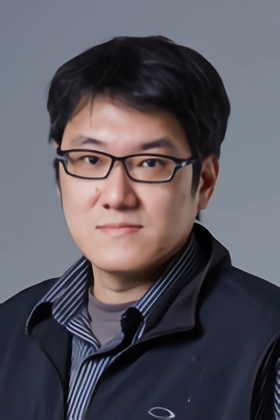}}]
{Yu Tsao} (Senior Member, IEEE) received his B.S. and M.S. degrees in Electrical Engineering from National Taiwan University, Taipei, Taiwan, in 1999 and 2001, respectively, and his Ph.D. degree in Electrical and Computer Engineering from the Georgia Institute of Technology, Atlanta, GA, USA, in 2008. From 2009 to 2011, he was a researcher at the National Institute of Information and Communications Technology, Kyoto, Japan, where he worked on research and product development in automatic speech recognition for multilingual speech-to-speech translation. He is currently a Research Fellow (Professor) and the Deputy Director of the Research Center for Information Technology Innovation at Academia Sinica, Taipei, Taiwan. He also serves as a Jointly Appointed Professor in the Department of Electrical Engineering at Chung Yuan Christian University, Taoyuan, Taiwan. His research interests include assistive oral communication technologies, audio coding, and bio-signal processing. Dr. Tsao is currently an Associate Editor for the IEEE/ACM TRANSACTIONS ON AUDIO, SPEECH, AND LANGUAGE PROCESSING and IEEE SIGNAL PROCESSING LETTERS. He was the recipient of the Academia Sinica Career Development Award in 2017, national innovation awards from 2018 to 2021, the Future Tech Breakthrough Award in 2019, the Outstanding Elite Award from the Chung Hwa Rotary Educational Foundation in 2019–2020, the NSTC FutureTech Award in 2022, and the NSTC Outstanding Research Award in 2023. He is the corresponding author of a paper that received the 2021 IEEE Signal Processing Society (SPS) Young Author Best Paper Award.
\end{IEEEbiography}

\end{document}